# Observing monomer - dimer transitions of neurotensin receptors 1 in single SMALPs by homoFRET and in an ABELtrap

André Dathe[a], Thomas Heitkamp[a], Iván Pérez[a], Hendrik Sielaff[a], Anika Westphal[b], Stefanie Reuter[b], Ralf Mrowka[b], Michael Börsch[a,c,*]

[a]Single-Molecule Microscopy Group, Jena University Hospital, Nonnenplan 2 - 4, 07743 Jena,
[b]Experimental Nephrology Group, Jena University Hospital, Nonnenplan 2 - 4, 07743 Jena
[c]Abbe Center of Photonics (ACP) Jena, Germany

## ABSTRACT

G protein-coupled receptors (GPCRs) are a large superfamily of membrane proteins that are activated by extracellular small molecules or photons. Neurotensin receptor 1 (NTSR1) is a GPCR that is activated by neurotensin, i.e. a 13 amino acid peptide. Binding of neurotensin induces conformational changes in the receptor that trigger the intracellular signaling processes. While recent single-molecule studies have reported a dynamic monomer – dimer equilibrium of NTSR1 *in vitro*, a biophysical characterization of the oligomerization status of NTSR1 in living mammalian cells is complicated. Here we report on the oligomerization state of the human NTSR1 tagged with mRuby3 by dissolving the plasma membranes of living HEK293T cells into 10 nm-sized soluble lipid nanoparticles by addition of styrene-maleic acid copolymers (SMALPs). Single SMALPs were analyzed one after another in solution by multi-parameter single molecule spectroscopy including brightness, fluorescence lifetime and anisotropy for homoFRET. Brightness analysis was improved using single SMALP detection in a confocal ABELtrap for extended observation times in solution. A bimodal brightness distribution indicated a significant fraction of dimeric NTSR1 in SMALPs or in the plasma membrane, respectively, before addition of neurotensin.

## 1 INTRODUCTION

G protein-coupled receptors (GPCRs) are the large superfamily of membrane proteins that are involved in controlling distinct physiological processes. GPCRs transmit the information received by absorbing photons, for example rhodopsin, or by binding small molecules from the extracellular side. The information is transferred across the plasma membrane barrier for intracellular signaling processes. GPCRs are important for neurotransmission, behavioral regulation, immune response, taste, smell or vision, respectively. Therefore they are targets for more than 30% of all therapeutic drugs applied today[1]. One central question regarding the first steps of signaling is the role of monomeric *versus* oligomeric GPCRs as the active transmitters.

We focus on the neurotensin receptor 1 (NTSR1) that belongs to the β group of class A GPCRs. NTSR1 participates in modulating dopaminergic systems, analgesia and inhibition of food intake in the brain and in the intestines[2, 3]. The 13 amino acid peptide neurotensin (ELYENKPRRPYIL) acts as the ligand for NTSR1 and binds with nanomolar affinity[4, 5]. Neurotensin binding causes conformational changes in NTSR1 that trigger signaling processes by G proteins inside the cells as well as interactions with kinases and arrestin molecules within minutes. The X-ray structure of NTSR1 with a bound ligand has been solved by the group of R. Grisshammer in 2012[6].

The possibility to express NTSR1 also in bacterial cells and to purify the receptor allows structural and functional biophysical studies on the single molecule level and in controlled membrane environments[7]. Recently transient dimerization of purified NTSR1 in a solid-supported artificial lipid bilayer system has been reported[8] by single-molecule FRET (smFRET) measurements between differently tagged receptors. We also have found a fraction of dimers of purified detergent-solubilized NTSR1[9] after reconstitution into liposomes at low concentrations[10].

........................................................................................................................

* email: michael.boersch@med.uni-jena.de; http://www.single-molecule-microscopy.uniklinikum-jena.de

Because the observed partial dimerization or oligomerization of reconstituted NTSR1 in the absence of its ligand neurotensin could result from a purification or from reconstitution artefacts, we started to investigate the oligomerization state of fluorescently tagged NTSR1 in its native lipid environment in living HEK293T cells. A mutant with mRuby3[11] as a fluorescent protein fused to the C-terminus of human NTSR1 was constructed. We applied widefield microscopy, structured illumination microscopy (SIM), fluorescence lifetime imaging microscopy (FLIM) and anisotropy imaging (*see* A. Westphal et al., Proc. SPIE, 10884, *in press* [2019]). However, our preliminary fluorescence correlation spectroscopy (FCS) measurements of NTSR1 in the plasma membranes indicated a concentration of the receptors that was too high to resolve individual NTSR1 molecules in the confocal detection volume.

A novel approach to dissolve native membranes with embedded proteins into nanometer-sized soluble patches without the use of any detergent requires an amphiphilic copolymer of styrene and maleic acid (see review[12]). Addition of this copolymer results in the formation of nanodiscs of the native lipids and proteins with a relatively uniform size of 10 nm[13]. The nanoparticles are negatively charged due to the maleic acid salt component. These small nanoparticles are often called SMALPs (styrene-maleic acid lipid particles) and comprise the native lipids without dilution. The native lipid mixture might control the activity of the GPCRs, and their homo- and hetero-oligomerization with other proteins[14, 15]. Here we used SMALPs to produce NTSR1 in native lipid membrane particles from living HEK293T cells and analyzed the oligomerization status of the receptor before addition of its ligand neurotensin.

Quantitative single nanoparticle spectroscopy in solution is facilitated by the extended observation times in an Anti-Brownian Electrokinetic trap (ABELtrap) as invented by A. E. Cohen and W. E Moerner[16-20]. We analyzed the brightness of the NTSR1-SMALPs in our ABELtrap[21-24] and found a bimodal intensity distribution indicating that NTSR1 likely exists as a monomer but also as a dimer or oligomer in the native membrane of mammalian cells before addition of neurotensin.

## 2 EXPERIMENTAL PROCEDURES

### 2.1 Stable expression of NTSR1-mRuby3 in living HEK293T cells

NTSR1 comprises seven transmembrane helices, an extracellular N-terminus and an intracellular C-terminus. At the C-terminus we inserted the fluorescent protein mRuby3[11]. The plasmid was introduced into human HEK293T FlpIn cells to generate a stable cell line for homogenous gene expression and protein production (for details see A. Westphal et al, Proc. SPIE 10884, *in press* [2019]). HEK293T FlpIn cells (ThermoFischer Scientific, Waltham, MA, USA) were cultured at 37°C in high-glucose Dulbecco's Modified Eagle Medium (DMEM, ThermoFischer Scientific, Waltham, MA, USA) supplemented with 10% fetal bovine serum (FBS) (Biochrom, Berlin, Germany) and 100 µg/ml hygromycin B (Roth, Karlsruhe, Germany). For spectroscopy soluble mRuby3 was expressed in *E. coli* and purified by Ni-NTA-based chromatography similar to published procedures[10].

### 2.2 Preparation of SMALPs

To prepare NTSR1-mRuby3 in SMALPs we used a large flask with an area of about 75 cm$^2$ for culturing of HEK293T FlpIn cells. After harvesting the cells by scraping and a brief centrifugation the cell pellet was quickly mixed with 1 ml SMA buffer (20 mM HEPES, 1 mM EGTA, 1 mM Mg-acetate, pH 7.4) containing 2% of the styrene-maleic acid copolymer (XIRAN SL25010-P20; Polyscope Polymers, Waalwijk, The Netherlands) and 5 U/ml freshly added benzonase. The solution was shaken for 1 h at 37 °C to induce SMALP formation. Ultracentrifugation (1 h, 100,000 x g, 4 °C) separated the soluble cell content including DNA fragments from a small pellet containing the SMALPs. The pellet was resuspended in 100 µl PBS buffer for further use.

### 2.3 Spectroscopy

Absorption spectra were measured in 100 µl quartz cuvettes (Hellma Analytics, Müllheim, Germany) using a Lambda 650 UV/Vis Spectrophotometer (PerkinElmer, Waltham, MA, USA). Fluorescence spectra and anisotropy spectra were measured using a PTI QuantaMaster 30 spectrofluorometer (HORIBA Scientific, Kyoto, Japan) with slit widths set to 5 nm and 1 s integration time per data point. Fluorescence lifetimes were measured using a modular fluorescence lifetime spectrometer FluoTime 100 (Picoquant, Berlin, Germany): Excitation with a 503 nm pulsed LED, controlled by a PDL 800-B driver, was set to a repetition rate of 10 MHz Fluorescence decay curves were recorded at 592 nm.

## 2.4 Microscopy of NTSR1-mRuny3 in HEK293T cells

To control stable expression of NTSR1-mRuby3 in living HEK293T cells widefield imaging and structured illumination microscopy (3D-SIM) were performed on a Nikon N-SIM / N-STORM microscope[25-28] using 561 nm laser excitation and the Nikon 60x water immersion objective with N.A. 1.27 (CFI Plan Apo IR 60x WI, Nikon, Germany) or the Nikon 100x silicon oil immersion objective with N.A. 1.35 (CFI SR HP lambda S 100XC Sil; temporary loan from Nikon, Germany), respectively. Images were recorded by a cooled EMCCD camera (iXon DU-897, Andor Technologies, Belfast, UK) using the Nikon "SIM561" optical filter sets. SIM images of living cells were recorded at 23 °C or at 37 °C. Nikon analysis software was used for SIM image reconstruction or deconvolution, and for 3D visualization using the maximum intensity projection option.

## 2.5 Confocal single-molecule microscopy setup

Our modular confocal setup for 4-channel single molecule spectroscopy in solution has been described before[29-33]. We used a 3D piezo scanning system (Physik Instrumente, Karlsruhe, Germany) for FLIM and anisotropy imaging mounted on an Olympus IX71 that was equipped with an 60x water immersion objective (UPLSAPO 60XW with N.A. 1.2, Olympus, Tokyo, Japan) or 60x silicon oil immersion objective (UPLSAPO 60XS2 with N.A. 1.3, Olympus) Here, ps-pulsed excitation was provided with 561 nm at 80.6 MHz (SC-450, Fianium, UK) using a tunable filter system (TuneBox with filter VersaChrome HC 617/14, AHF, Tübingen, Germany) combined with an additional laser clean-up filter centered at 561 nm with 4 nm FWHM (AHF) and a polarizing beam splitter cube (Thorlabs, Newton, NJ, USA) for linear polarization. Photons were detected by two of the to four single-photon counting avalanche photodiodes (i.e. two SPCM-AQRH-14-TR APDs optimized for time-resolved measurements, Excelitas Technologies GmbH, Wiesbaden, Germany) for lifetime and anisotropy easurements[31]. One TCSPC card (SPC-150, Becker&Hickl, Berlin, Germany) in combination with an 8-channel router (HRT-82, Becker&Hickl) recorded the photons. Precise timing of the two APDs for FLIM and time-resolved anisotropy measurements required a careful shifting of the photon signals of one APD with ps time resolution that was achieved by an electronic ps delay generator ($PSD-065-A-MOD, Micro Photon Devices, Bolzano, Italy). A second ps delay generator war used to amplify the trigger signals from the Fianium laser and to synchronize the TCSPC electronics. FLIM and time-resolved anisotropy images were analyzed using the SPCImage software (Becker&Hickl), whereas intensity-based anisotropy images were analyzed using counter cards (National Instruments, Austin, TX. USA) and custom Matlab scripts (Mathworks, Natick, MA, USA)[34-39].

## 2.6 Confocal ABELtrap setup

For the SMALP measurements with NTSR1-mRuby3 in our ABELtrap[21-24], the 491 nm laser (Calypso; Cobolt, Hübner Photonics, Kassel, Germany) was replaced by a new combined 4-line laser operated at 561 nm (Skyra, Cobolt; a temporary loan from Cobolt, Solna, Sweden, and initiated by von Gegerfelt Photonics, Bensheim, Germany). The schematic of the ABELtrap setup is shown in Figure 1 below.

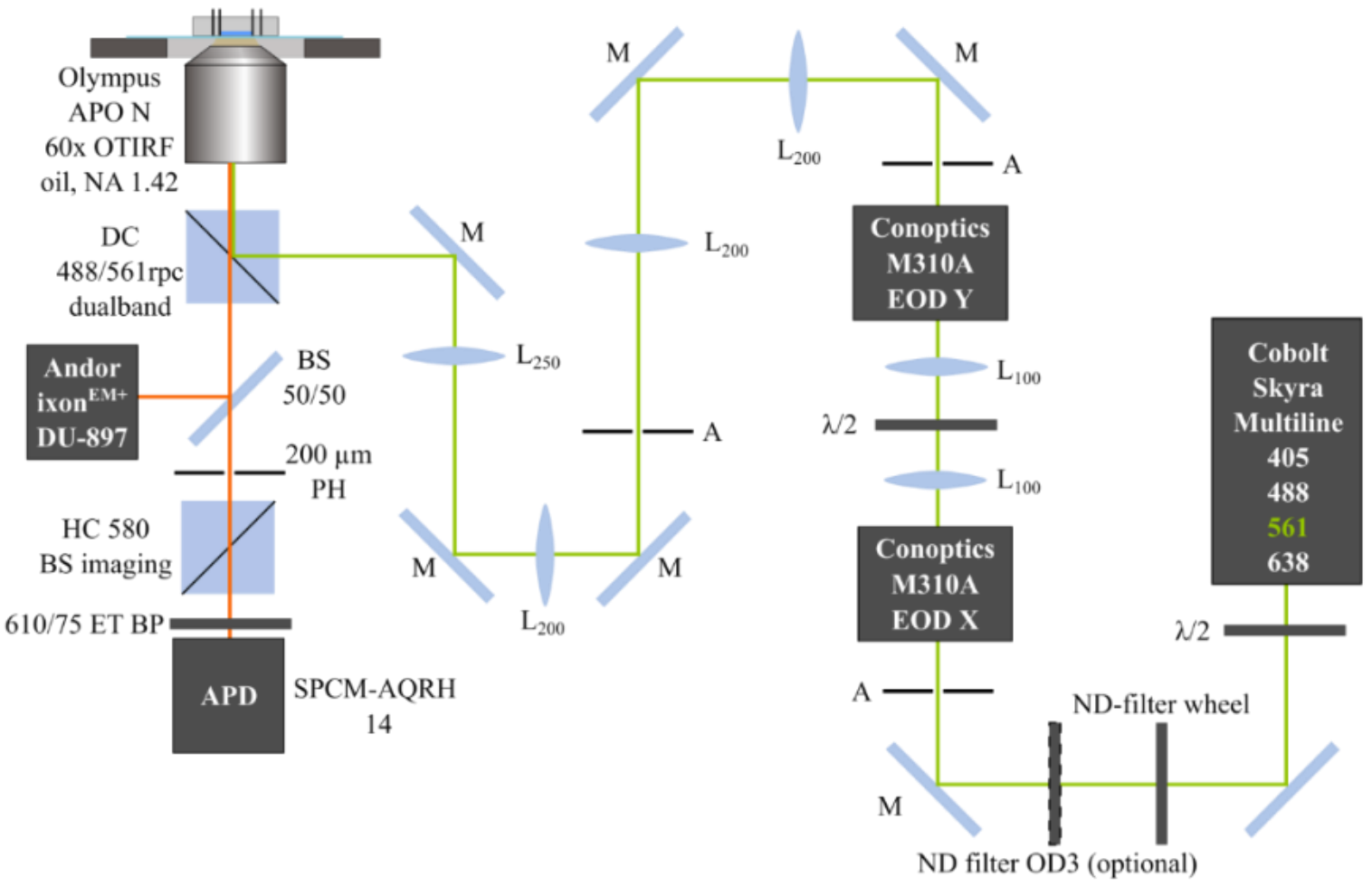

**Figure 1**: Setup of the ABELtrap using a 4-line continuous-wave laser for excitation (Skyra, Cobolt).

The laser focus in the ABELtrap was moved by two EODs (M310A, Conoptics, Danbury, CT, USA) and set to a 32 point knight pattern[40] at 5 kHz per pattern. Laser power was set to 40 µW in the focal plane. Photons were detected by one APD (SPCM-AQRH-14, Excelitas Technologies) and recorded by a FPGA card (7852R, National Instruments, ) and a TCSPC card (DPC230, Becker&Hickl) in parallel. electrode feedback voltages were limited to ±10 V. Disposable microfluidic chambers for the ABELtrap were built from plasma-cleaned cover glass and PDMS (Sylgard 184 elastomer, Dow, Midland, MI, USA). PDMS with a ratio of 1:10 from polymer:hardener was hardened for 4 h at 70°C. The PDMS chambers were filled with 10 µl of the analyte solutions.

# 3 RESULTS

## 3.1 Spectroscopic characterization of soluble mRuby3

To enable future FRET measurements of conformational changes in NTSR1, we have constructed a receptor mutant comprising a tetracysteine motif for fluorescein-based FlAsH labeling as the FRET donor and a possible FRET acceptor mRuby3 fused to the C-terminus of NTSR1 similar to other GPCR mutants[41]. According to published data[11], mRuby3 has a moderate fluorescence quantum yield of 45% and can be excited with the 561 nm laser line. We expressed mRuby3 in *Escherichia coli* bacteria and purified the protein. Figure 2 shows absorbance, fluorescence excitation and emission spectra of soluble mRuby3 in buffer. The fluorescence maxima were determined to 555 nm for excitation and 594 nm for emission, respectively, and the absorbance maximum was 558 nm.

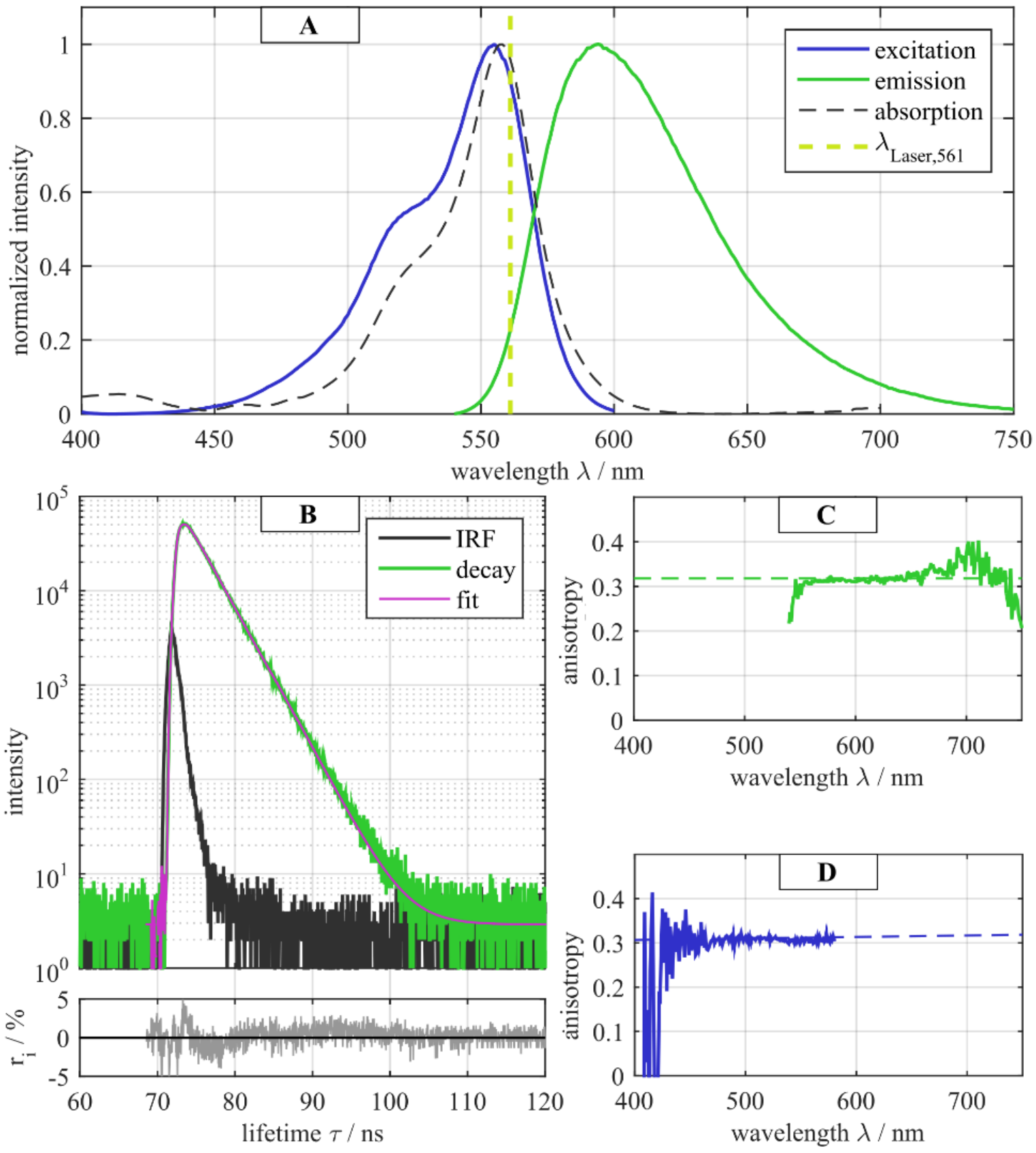

**Figure 2**: **A**, Absorbance (dashed grey), fluorescence excitation (blue) and emission (green) spectra of soluble mRuby3 in buffer, with absorbance maximum at 558 nm, excitation maximum at 555 nm and emission maximum at 594 nm. The vertical line indicates excitation with the 561 nm. **B**, fluorescence decay histogram (green) with an average fluorescence lifetime $\tau$=2.85 ± 0.01 ns (monoexponential fitting in purple) and instrument response function of the pulsed 503 nm LED (IRF in black). Residuals $r_i$ (shown in grey, lower panel) resulted in a $\chi^2$ = 1.48. **C, D,** fluorescence anisotropy spectra with emission anisotropy spectrum (**C**) in green and excitation anisotropy spectrum (**D**) in blue. Dashed lines are mean anisotropies of $r_{Ex}$=0.32 ± 0.02 and $r_{Em}$=0.31 ±0.01, respectively.

The fluorescence lifetime of soluble mRuby3 was measured using a pulsed 503 nm LED and recorded at the emission maximum of 592 nm. Mono-exponential decay fitting yielded $\tau = 2.85 \pm 0.01$ ns. Multi-exponential decay fitting did not improve the residual distribution significantly. Fluorescence anisotropy spectra revealed the expected high static anisotropies of $r_{Ex} = 0.32 \pm 0.02$ (excitation) or $r_{Em} = 0.31 \pm 0.01$ (emission) for a fluorescent protein that contains a rigid embedding of the chromophore within the β-barrel structure of the protein.

### 3.2 Diffusion and brightness of soluble mRuby3 and SMALPs with NTSR1-mRuby3

Next we measured the diffusion and brightness properties of soluble mRuby3 in a confocal microscope for comparison with organic fluorophores as references and with NTSR1-mRuby in SMALPs. The autocorrelation functions shown in Fig. 3 were recorded using the ABELtrap setup (but without trapping in PDMS microfluidics) with 561 nm cw laser excitation at 100 µW.

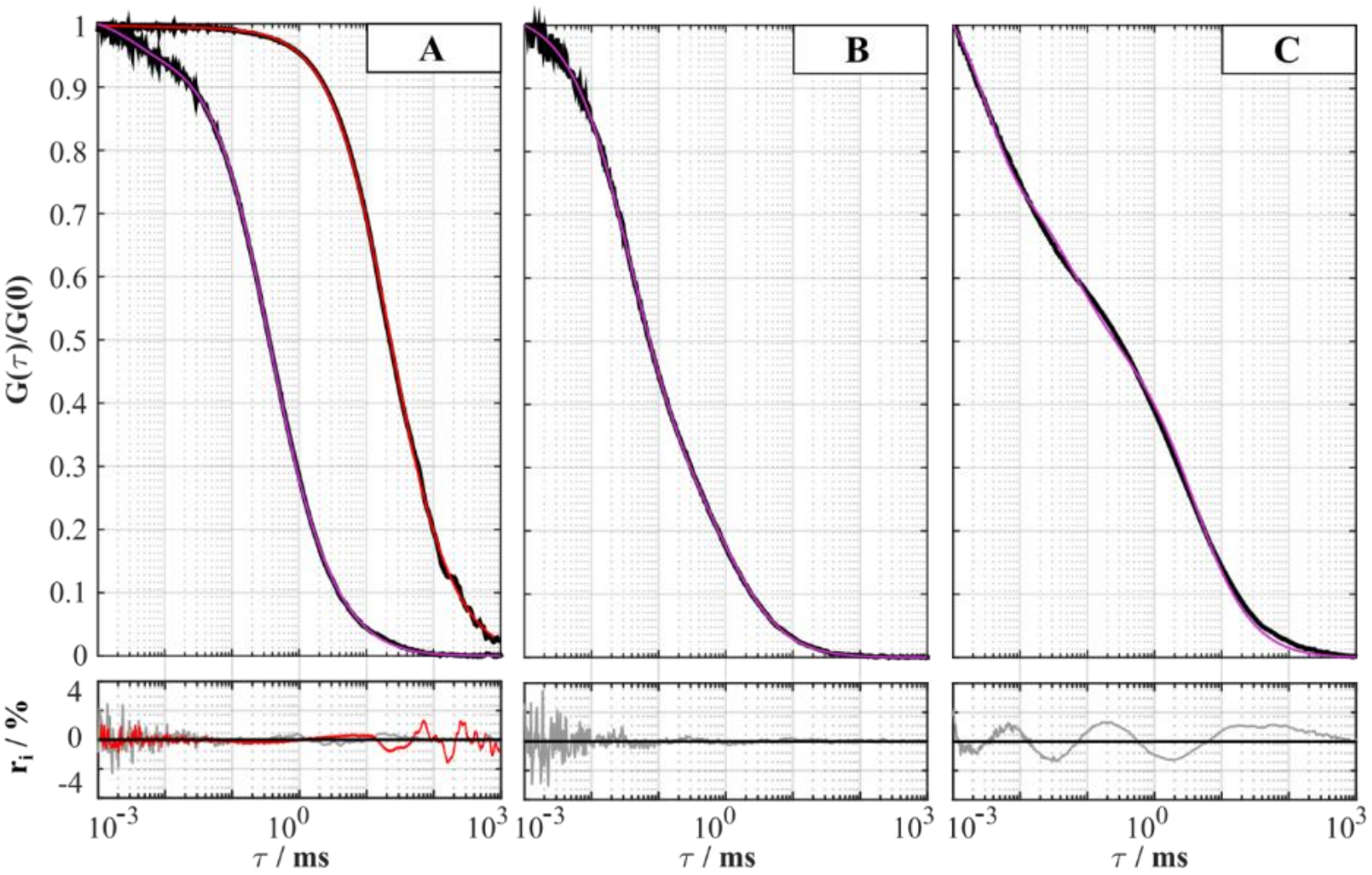

**Figure 3**: Normalized autocorrelation functions in black (FCS) and fitting (purple or red curves) according to given models. **A**, Atto565-maleimide and 0.1 µm fluorescent beads (TetraSpeck) in $H_2O$. **B**, soluble mRuby3 in buffer. **C**, NTSR1-mRuby3 in SMALPS in buffer (see text below for details).

Fluorescent beads in $H_2O$ with 0.1 µm diameter and a known diffusion coefficient D = 4.4 µm²/s served as a reference[42] to determine the dimensions of the confocal detection volume. The average diffusion time of the beads was $\tau_{Diff} = 21.7$ ms (Fig. 3 A). The confocal radius (x, y) was determined to $\omega_0 = 0.66$ µm according to the relation of diffusion time and diffusion coefficient $\tau_{Diff} = \omega_0^2/4D$ for isotropic Brownian motion. For Atto565-maleimide reacted with mercaptoethanol and diluted in $H_2O$ we calculated D =320 µm²/s from a diffusion time of $\tau_{Diff} = 340$ µs. A triplet lifetime for Atto565-maleimide was found with $\tau_T = 3$ µs.

When soluble mRuby3 was diluted to ~0.1 nM in buffer (10 mM Tris, 0.5 mM EGTA, 0.5 mM Mg-acetate, pH 7.4) and added to the cover glass as a 30 µl droplet for FCS measurements, we noticed a fast binding of the protein to the glass surface. Using other buffers or pure $H_2O$ did not prevent sticking. Accordingly any FCS results for diffusion and brightness properties might be flawed[30]. Figure 3 B shows the autocorrelation function of mRuby3. Fitting the FCS was possible only after adding a second diffusion term or a bunching term, respectively, for an additional dark state of mRuby3. The longer diffusion time $\tau_{Diff} = 980$ µs corresponded to D = 98 µm²/s (with a relative amplitude of 32%), i.e. in agreement with the diffusion coefficient D = 87 µm²/s of GFP[43] with similar size and shape. The dark state of mRuby3 exhibited a correlation time of ~40 µs, in addition to a short 3 µs triplet state lifetime.

Comparing intensities in the fluorescence spectra of NTSR1-mRuby3 in SMALPs and soluble mRuby3 with known concentration yielded an estimated NRSR1-mRuby3 in SMALP concentration of about 1 nM. The diffusion time $\tau_{Diff} = 5.88$ ms was measured by FCS and corresponded to D = 16 µm²/s (Fig. 3 C), i.e. 20 times longer than for Atto565-maleimide in $H_2O$. With a hydrodynamic radius of rhodamines like Atto565 in the range of 1 nm, the hydrodynamic radius for the SMALPs was significantly larger than the expected 10 nm diameter. However, as seen in the residual distribution in Fig. 3 C, this FCS fitting was not perfect. Systematic deviations of the residuals might indicate a broadened distribution of SMALP sizes. The additional dark state of NTSR1-mRuby3 was fitted with 90 µs.

### 3.3 Single-molecule lifetimes and anisotropies of NTSR1-mRuby3 in SMALPs in solution

Next we investigated NTSR1-mRuby3 in single SMALPs freely diffusing in solution using a different confocal setup. Here we applied linearly polarized, pulsed excitation with 561 nm to measure fluorescence lifetime and anisotropy (or homoFRET, respectively) in individual photon bursts. Figure 4 shows seven examples of photon bursts.

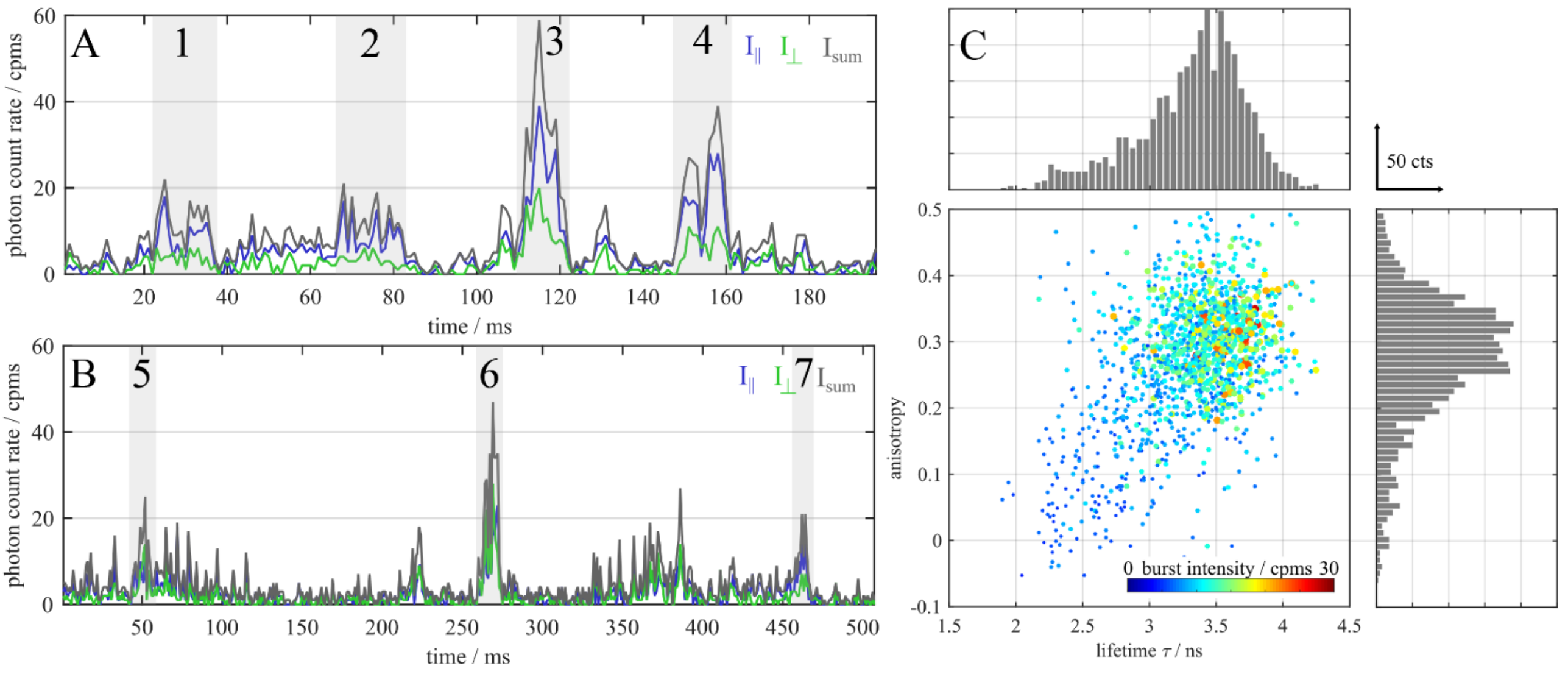

**Figure 4**: **A, B**, time traces with photon bursts of single SMALPs in solution. Linearly polarized, pulsed excitation yielded anisotropies per burst as calculated from intensities recorded with parallel (blue traces) or perpendicular (green traces) polarization as well as mean burst intensities from the sum of both (grey traces). Parameters of the marked bursts are discussed in the text. **C**, distribution of brightness-dependent lifetimes and associated anisotropies of individual SMALPs.

Marked by grey bars in Fig. 4 A, photon burst 1 exhibited an anisotropy value $r = 0.38$ with peak counts of ~20 per ms (cpms), photon burst 2 exhibited $r = 0.46$ with similar peak counts, photon burst 3 exhibited $r = 0.27$ but ~60 cpms as peak counts, and photon burst 4 exhibited $r = 0.42$ with 40 cpms as peak count. While these anisotropies were similar for these four examples, other anisotropies were found as well. Photon burst 5 in Fig. 4 B exhibited $r = 0.1$ with a peak count of 20 cpms, photon burst 6 exhibited $r = 0.01$ with a peak count of 50 cpms, and photon burst 7 exhibited $r = 0.3$ with a peak count of 20 cpms. For all photon bursts the fluorescence lifetime was calculated from the sum intensity (grey traces). $\tau$ was mostly found between 3 ns an 4 ns. Figure 4 C summarizes lifetimes and associated anisotropies of the short photon bursts detected while traversing the confocal detection volume within 10 ms to 100 ms. Anisotropies centered around $r = 0.3$ and lifetimes around $\tau = 3.5$ ns. Lower mean intensities resulted in a minor population with shortened lifetimes between 2 ns and 3 ns and associated smaller anisotropies $0 < r < 0.3$. However, the low photon numbers in these short SMALP photon bursts did not provide precise lifetimes nor anisotropies and might have caused the broad distribution of values in Fig. 4 C.

In addition we calculated the autocorrelation function for the SMALPs and found diffusion times to be 10 to 15 times longer than for Atto565-maleimide. Importantly, the dark state of mRuby3 had the same correlation time around 50 µs despite the different size of the confocal volume in this setup. This supported an interpretation as a photophysical state and not a second diffusing species. We also compared the mean brightness calculated from the average fluorescence intensity of the time trace divided by the mean number of fluorescent molecules in the confocal volume. Single Atto565-maleimide had a brightness of 15.4 kHz (or cpms), single soluble mRuby3 had a brightness of about 4 kHz, and single

SMALPs had a brightness of 11 kHz under the same experimental conditions (pulsed excitation at 561 nm with 150 µW, 80.6 MHz repetition rate, enlarged confocal volume with $\tau_{Diff}$ = 700 µs for Atto565-maleimide in $H_2O$).

### 3.4 Brightness of individual NTSR1-mRuby3 in SMALPs hold in solution by the ABELtrap

The distinct mean brightness of soluble mRuby3 and NTSR1-mRuby3 in SMALPs required detailed brightness analysis with extended observation times of single SMALPs as provided by the confocal ABELtrap. SMALPs were efficiently trapped when the target diffusion coefficient was set to D = 35 µm²/s and the initial electrokinetic mobility ranging between |µ| = 300 - 500 µm$(Vs)^{-1}$ in the Kalman filter-based LABVIEW trapping software[20, 21]. The excitation power was set to 40 µW at 561 nm by the Skyra multiline laser. Figure 5 shows six examples of SMALPs that were trapped for 300 ms to 500 ms, i.e. were observable about 50 to 100 times longer than freely diffusing SMALPs.

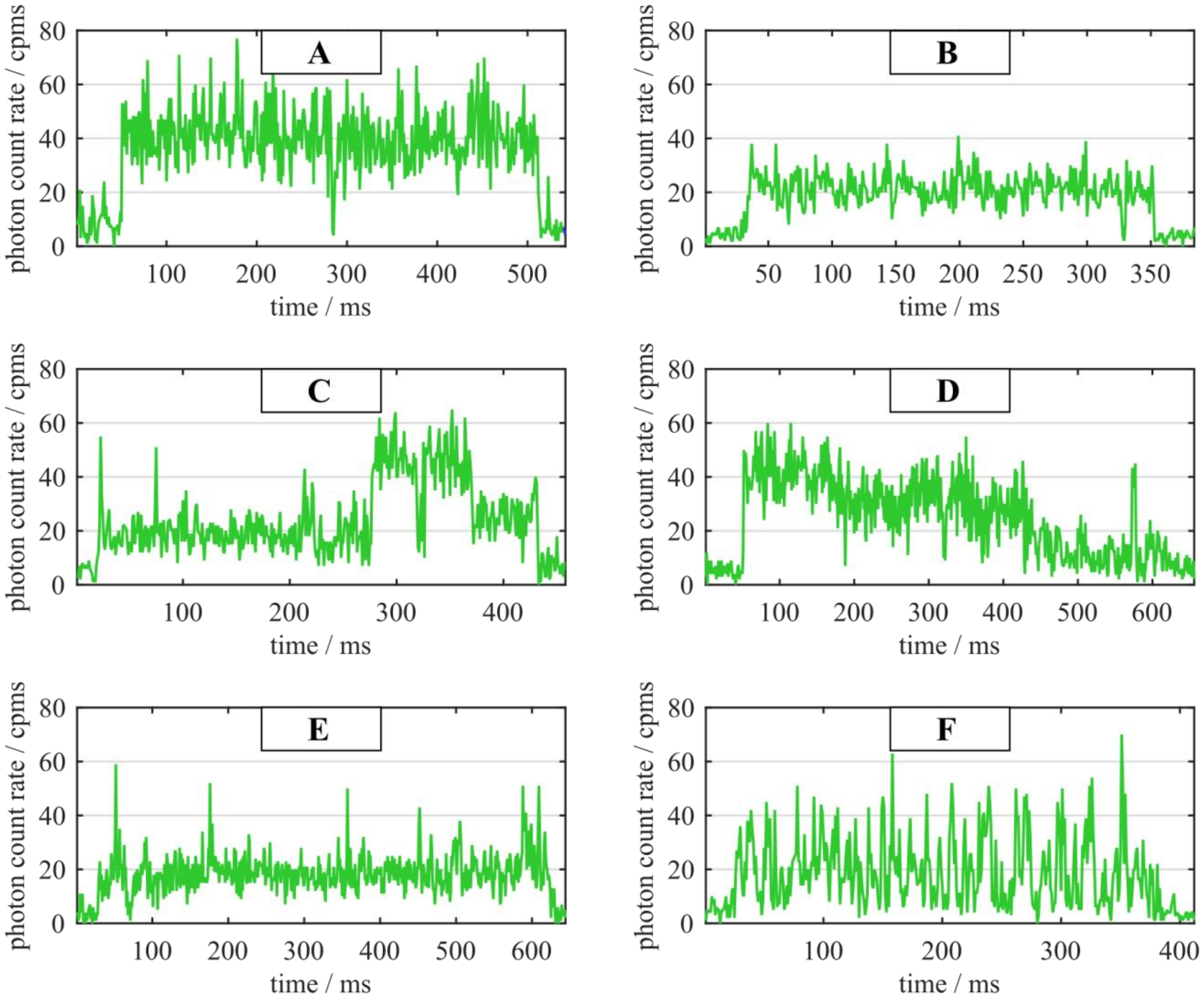

**Figure 5**: **A-F**, six photon bursts of SMALPs hold the ABELtrap. **A, B,** the mean intensities were constant but varied; **C**, the intensity switches between two states within the photon burst; **D**, apparent stepwise intensity loss within the photon burst; **E**, short switching to high intensities; **F**, strong and fast blinking within the photon burst.

The mean photon count rates of the trapped SMALPs were in photon burst A 36.1 cpms, in B 18.6 cpms, in C 23.9 cpms, in D 24.2 cpms, in E 18.1 cpms, and in F 17.9 cpms. We manually removed photon burst with strong intensity fluctuations (see Fig. 5 F) and assigned different intensity levels within a photon burst when necessary (see Fig. 5 C). The histogram of all intensity states of trapped SMALPs is shown in Fig. 6 A. The bimodal distribution was fitted by two Gaussians yielding a lower mean $\mu_1$= 15.4 cpms with $\sigma_1$= 4.9 cpms and a higher mean $\mu_2$ = 33.7 cpms with $\sigma_2$ = 6.2 cpms. The distribution of trapping times of SMALPs is shown in Fig. 6 B indicating that most of the SMALPs were trapped for less than 500 ms, but a few SMALPs were trapped for up to 2 s.

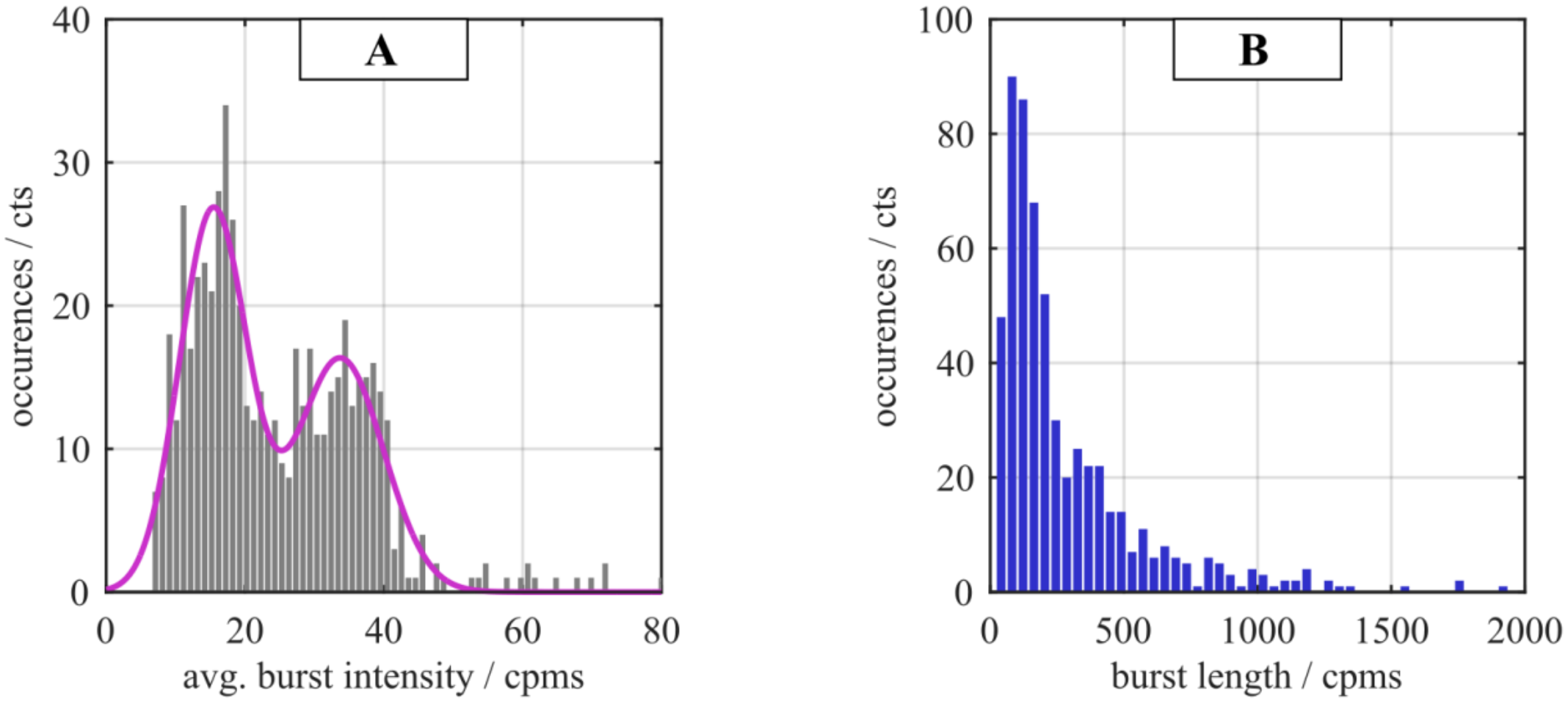

**Figure 6**: **A**, distribution of mean photon burst intensity levels found for single SMALPs containing NTSR1-mRuby3 as measured in the ABELtrap. Two distinct populations of intensity levels were observed from a total of 580 photon bursts. A bimodal Gaussian distribution was fitted and mean positions of $\mu_1$ = 15.4 cpms and $\mu_2$ = 33.7 cpms determined. **B**, distribution of photon burst durations of the 580 trapped SMALPs.

## 4 DISCUSSION

A biophysical discrimination of monomers *versus* dimers or oligomers of membrane receptors like GPCRs in living cells is possible using precise distance measurements between these proteins. FRET could be employed to measure distances below 10 nm[44]. Therefore the GPCRs have to be tagged with one in two distinct fluorophores, for example genetically encoded fluorescent proteins. If FRET is detected in pixels of a microscopic image, at least a fraction of GPCRs exists as a dimer or oligomer. However, if the number of receptors per membrane area is too high, a spatial localization of monomers and dimers becomes difficult or even impossible. To generate a FRET pair with distinguishable spectra a plasmid has to be constructed that enables the expression of the two types of GPCRs at identical or at least similar concentrations, and with similar maturation times of the fluorescent proteins. Using only one type of tagged GPCR for homoFRET is less complicated. As the readout for homoFRET the fluorescence anisotropy is analyzed[45].

Optical diffraction defines the minimal detection volume in microscopy and sets an upper limit for receptor numbers to be discriminated within the detection volume. To circumvent an ensemble averaging of monomers, dimer and oligomers of GPCRs within the detection volume, the membrane has to be subdivided into 10 nm patches. Then, in each of the patches would likely exist either a monomer, a dimer, or an oligomer. One way to dissolve the membranes without adding detergent is the use of styrene-maleic acid copolymers that can form 10 nm-sized SMALPs directly from living cells. We have implemented the protocols[14, 15] to produce SMALPs that contain all membrane proteins of HEK293T cells including the neurotensin receptor 1 that was tagged at the C-terminus with mRuby3. The diffusion times of these SMAPLs were found to correspond approximately to the expected 10 to 15 nm diameter.[12]

Confocal measurements of the freely diffusing SMALPs showed photon bursts with ms duration. Therefore the number of photons in each photon burst were small. Accordingly calculations of fluorescence lifetimes and fluorescence anisotropy for the SMALPs were flawed. Lifetime information was used to identify the fluorescent protein mRuby3 given an excitation at 561 nm and emission detected between 575 nm and 645 nm, and anisotropy was used to detect homoFRET. The 2D distribution of lifetime and associated anisotropy values indicated that a small fraction of NTSR1-mRuby3 in SMALPs might exist as a dimer, i.e. in the membranes of living HEK293T cells already before activation of the receptor with neurotensin.

Longer observation times of individual SMALPs were required to provide stronger evidence for a fraction of NTSR1 dimers in the absence of neurotensin. The ABELtrap was invented by A. E. Cohen and W. E. Moerner and is the method of choice to hold single molecules and nanoparticles in solution in place and to record all fluorescence parameters[18, 19, 46-49]. Here we utilizes an ABELtrap with cw laser excitation at 561 nm to determine the brightness levels of NTSR1-mRuby3 in SMALPs. The SMALPs were trapped in solution for up to 2 s. Distinct brightness levels were identified that remained constant throughout the photon burst. Some SMALPs exhibited transitions between brightness levels including single-molecule on/off blinking behavior. For the assigned stable intensity levels we found a bimodal intensity distribution. Because the maxima of the two populations differed by approximately a factor of 2, we interpret this distribution as a strong indication for the coexistence of monomeric and dimeric NTSR1 in the plasma membrane of human HEK293T cells even in the absence of its agonist neurotensin.

To support this conclusion we will extend the fluorescence detection possibilities of our ABELtrap by adding a second APD for homoFRET or anisotropy recording, and by applying pulsed lasers for lifetime measurements simultaneously. Thus we will improve the identification of the fluorescent tag on NTSR1 and discriminate SMALPs comprising other membrane proteins.

**Acknowledgements**

The authors thank all members of our research groups who participated in genetics, biochemistry and cell culture, and especially thank M. Dienerowitz, B. Su and N. Zarrabi who assembled our version of the ABELtrap. We are grateful for the loan of the Skyra multiline laser (Cobolt, Hübner Photonics; initiated by von Gegerfelt Photonics). Financial support for the Nikon N-SIM / N-STORM superresolution microscope by the State of Thuringia (grant FKZ 12026-515 to M.B.), for A.W. through LOM funds 2017 by the Jena University Hospital (to M.B.), for H.S. and T.H. in part by the Deutsche Forschungsgemeinschaft DFG in the Collaborative Research Center/Transregio 166 “ReceptorLight” (project A1 to M.B.) is gratefully acknowledged. The ABELtrap was realized by additional DFG funds (grants BO1891/10-2, BO1891/15-1, BO1891/16-1, BO1891/18-2 to M.B.) and was supported by an ACP Explore project (M.B. together with J. Limpert) within the ProExcellence initiative ACP2020 from the State of Thuringia.